\documentclass[journal,11pt,onecolumn,draftclsnofoot]{article}

\usepackage{amsmath}
\usepackage{float}
\usepackage{braket}

\usepackage{ mathrsfs }
\usepackage{graphicx}

\usepackage{mathrsfs}  
\usepackage{siunitx}

\newcommand{\refEq}[1] {Eq. (\ref{#1})}

\newcommand{\refFig}[1] {Fig. \ref{#1}}

\usepackage{lineno}

\usepackage{subcaption}
        \usepackage{setspace}

\usepackage[margin=1in]{geometry} 

\begin{document}

\title{\bf High-efficiency microwave-optical quantum transduction based on a cavity electro-optic superconducting system with long coherence time}

\author{Changqing Wang$^1$ \and Ivan Gonin$^1$ \and Anna Grassellino$^1$ \and Sergey Kazakov$^1$ \and Alexander Romanenko$^1$ \and Vyacheslav P Yakovlev$^1$ \and Silvia Zorzetti$^{1,2}$}

\date{
	$^1$Fermi National Accelerator Laboratory, Batavia, IL, USA \\$^2$Corresponding author: zorzetti@fnal.gov\\%
}

\maketitle


\section*{\bf Abstract}
Frequency conversion between microwave and optical photons is a key enabling technology to create links between superconducting quantum processors and to realize distributed quantum networks. We propose a microwave-optical transduction platform based on long-coherence-time superconducting radio-frequency (SRF) cavities coupled to electro-optic optical cavities to mitigate the loss mechanisms that limit the attainment of high conversion efficiency. 
In the design, we optimize the microwave-optical field overlap and optical coupling losses, while achieving long microwave and optical photon lifetime at milli-Kelvin temperatures. This represents a significant enhancement of the transduction efficiency up to 50\% under pump power of \SI{140}{\micro\watt}, corresponding to few-photon quantum regime. Furthermore, this scheme exhibits high resolution for optically reading out the dispersive shift induced by a superconducting transmon qubit coupled to the SRF cavity. We also show that the fidelity of heralded entanglement generation between two remote quantum systems is enhanced by the low microwave losses. Finally, high-precision in quantum sensing can be reached below the standard quantum limit. 


\section*{\bf Introduction}
In the past decade, quantum information science and technology has been greatly boosted by the advancement of superconducting circuits \cite{schoelkopf2008wiring,wendin2017quantum,blais2021circuit}. However, the cryogenic temperature requirement restricts the capability to transfer quantum information between local quantum processors and therefore to build large-scale quantum networks \cite{pirandola2016physics}. Optical photons with frequency of hundreds of Tera-Hertz (\si{\tera\hertz}) are immune to thermal noise at room temperature and can be transferred distantly via the free-space and fiber communication technologies. As a result, optics has shown great promises for quantum communication and for connecting distant quantum nodes \cite{chen2021integrated}. Quantum transducers are required to mediate the huge energy gap between microwave and optical photons, about five orders of magnitude, and therefore to integrate optical communication systems with superconducting quantum devices \cite{lauk2020perspectives,han2021microwave}. 

Past years have witnessed extensive explorations of microwave-optical quantum transduction based on various kinds of hybrid platforms, including electro-optomechanics \cite{andrews2014bidirectional,arnold2020converting}, magnonics \cite{zhang2016optomagnonic,hisatomi2016bidirectional,zhang2014strongly,zhu2020waveguide}, piezo-optomechanics \cite{mirhosseini2020superconducting,jiang2020efficient,forsch2020microwave,han2020cavity},  atoms \cite{bartholomew2020chip,adwaith2019coherent,vogt2019efficient}, rare-earth ions \cite{o2014interfacing, fernandez2015coherent}, electro-optic nonlinear materials \cite{holzgrafe2020cavity,javerzac2016chip, fu2021cavity,xu2021bidirectional,mckenna2020cryogenic,soltani2017efficient}, and others. Among them, the electro-optic effect provides the most direct way to couple microwave and optical fields. The optical refractive index is modulated by the microwave electric field through the second-order electro-optic nonlinearity ($\chi^{(2)}$) \cite{tsang2010cavity,tsang2011cavity}. As a result, frequency conversion has been demonstrated based on microwave cavities integrated with nonlinear optical resonators made of electro-optic materials such as Lithium niobate (LN) \cite{wang2018integrated} and Aliminum nitride (AlN) \cite{fan2018superconducting}. In particular, the advancement of thin-film LN enables the fabrication of LN optical resonators with ultrahigh optical quality \cite{zhang2017monolithic}, offering a promising platform for microwave-optical frequency conversion \cite{javerzac2016chip, fu2021cavity,xu2021bidirectional,mckenna2020cryogenic,soltani2017efficient}. Yet the small mode volume leads to the overheating of the devices when the pump power applied to the optical resonator is high. Thus the trade-off between the thermal noise \cite{sahu2022quantum} and transduction efficiency makes it hard to realize a high-efficiency quantum transducer operating near the quantum threshold. Not long ago, three-dimensional cavities were exploited for transduction with bulk LN crystal optical resonators embedded, where the large mode volume gives rise to larger pump power tolerance \cite{hease2020bidirectional, rueda2019electro, sahu2022quantum,rueda2016efficient}. Pioneer work shows
\SI{0.03}{}\% photon conversion efficiency, and \SI{1.1}{} added output noise photons per second at \SI{1.48}{\milli\watt} pump power \cite{hease2020bidirectional}, demonstrating how three-dimensional geometries are transformative in quantum transduction, reducing the detrimental effect of thermally induced quasiparticles in the frequency conversion process. So far, the transduction efficiency is limited by the low quality factor (Q) of the microwave resonators, as well as by the non-optimal alignment between the microwave and optical mode distributions.

Recently, a new horizon has arisen for overcoming quantum decoherence with three-dimensional bulk Niobium superconducting cavities \cite{romanenko2020three, berlin2022searches,alam2022quantum,romanenko2017understanding}. Such cavities developed at Fermilab for particle accelerators exhibit record-long photon lifetime at milli-Kelvin (\SI{}{\milli K}) temperature, and thus mitigate the loss of microwave photons and greatly enhance light-matter interaction in the quantum regime. Moreover, the flexibility and variety of cavities' geometries provides full degree of freedom for mode selection and mode profile engineering. In addition, the large vacuum space in the mode volume allows the integration with various other materials for achieving hybrid quantum systems with rich functionalities.  

In this paper, we propose a high-efficiency, low-noise quantum transducer based on a three-dimensional architecture, with a bulk Niobium superconducting cavity integrated with a LN optical resonator. We maximize the microwave-optical interaction by careful RF design and optimization of the overlap between the optical field in the LN crystal structure and microwave field. We study and optimize the optical coupling strength between the prism coupler and the optical resonator through multi-physics simulations. We demonstrate through simulations that our system can achieve 50\% quantum transduction efficiency at a \SI{140}{\micro\watt} pump power, with an operating bandwidth of \SI{100}{\kilo\hertz}. Furthermore, with a superconducting qubit coupled to the microwave cavity, the transducer can readout the quantum state of the qubit with high resolution of the dispersive shift. Finally, such transducer device can be exploited for high-fidelity heralded quantum entanglement generation between two distant quantum processing units, as well as for high-precision microwave signal measurement and quantum sensing. 

\section*{\bf Results}
\subsection*{Device design}
\begin{figure}[!htb] 
	\centering
	\includegraphics[width=0.45\textwidth]{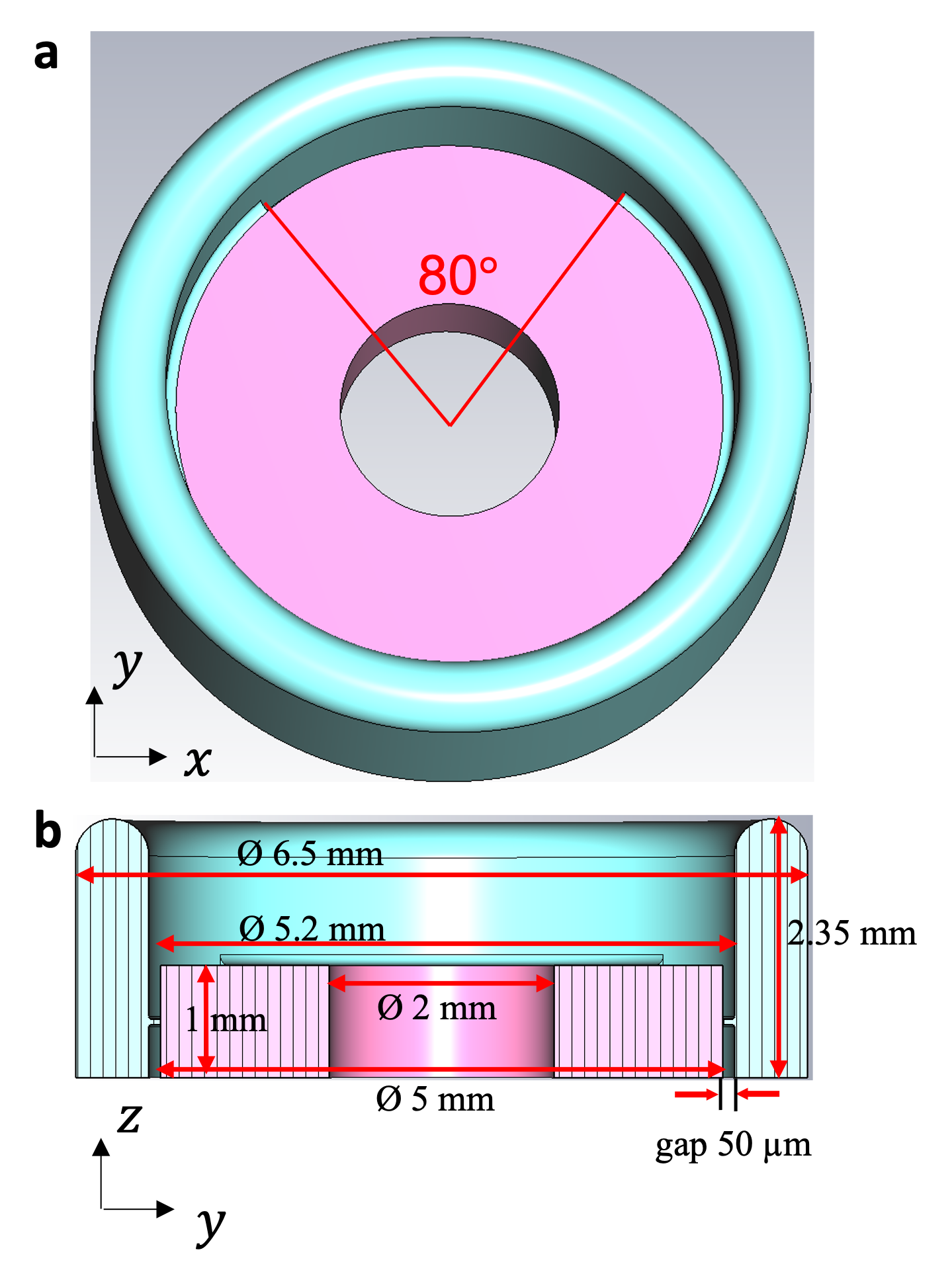}
	\caption{\normalsize Superconducting RF (SRF) cavity design for quantum transduction. \textbf{a} Top view of the superconducting microwave cavity enclosing a bulk Lithium Niobate (LN) optical resonator. \textbf{b} Side view of the cross section of the RF cavity with the enclosed LN resonator.} 
	\label{Fig1}
\end{figure}

The system we propose consists of a LN whispering galley mode (WGM) optical resonator enclosed in a 3D bulk niobium superconducting RF cavity (\refFig{Fig1}). The LN resonator with a \SI{5}{\milli\meter} diameter is made of a disk with the central part of the disk removed, and has the crystal axis in the z direction (z-cut), aligned to the electric field distribution. The electric field is enhanced by defining two symmetric interaction sections with a narrow lateral vacuum gap of \SI{50}{\micro\meter} spanning \SI{80}{\degree} in the azimuthal direction. Such a design reshapes the electric field distribution and enables the concentration of the microwave field onto the region near the rim of the LN disk where the WGM is located. This is the key to enhancing the interaction between the microwave and optical modes. It is noted that previous designs using electrodes deposition on the LN crystal to concentrate the microwave electric field, requires more complicated fabrication techniques including metal evaporation and photolithography, which may cause additional optical losses to the WGM. In contrast, our design is purely based on engineering of the LN crystal embedded in the microwave cavity. 
The diamond prism for the optical coupling is placed in the area with a larger gap (H up to \SI{1}{\milli\meter}).

Bulk Nb SRF cavities are attractive for quantum applications because of the high density of
the electromagnetic fields in large microwave volumes. SRF cavities built for particles accelerators have demonstrated $E_{max}=$\SI{10}{\mega\volt/\meter}, with $Q > 10^{10}$ \cite{romanenko2020three}. In transduction, the high field density within the cavity enhances the light-matter interactions and thus the electro-optic effect. 

The LN crystal enclosed in the SRF cavity occupies most of the volume, with a participation ratio ($p$) in the order of 0.96 (96\%). The microwave quality factor is mostly determined by the losses in the dielectric crystal, several other sources of losses are also present at cryogenic temperatures including the two-level-system (TLS) loss of impurities in the LN crystal, the piezo-electric loss, the thermal quasiparticle loss, and others \cite{mcrae2020materials}. For simplicity, we use a single value of loss tangent ($\tan\delta$) including all channels of losses in our evaluation, which has been reported to be in the order of $10^{-5}$ at cryogenic temperatures \cite{goryachev2015single}. The participation ratio and the quality factor are described as follow.

\begin{equation}
p=\frac{\int_{V_{d}} \epsilon_d |E|^2 dV_d}{\int_V\epsilon_0|E|^2dV},
\end{equation}

\begin{equation}
    Q=\frac{\omega_0U}{P_{loss}}=\frac{\omega_0\mu_0\int_V |H|^2dV}{\int_S|H|^2 dS},
\end{equation}
where $V$ is the total volume, and $V_d$ is the volume of the dielectric crystal. $\epsilon_{d,0}$ are the dielectric permittivities of the dielectric LN crystal and vacuum, respectively. The quality factor ($Q$) is the ratio between the energy stored in the cavity ($U$) and the dissipated power ($P_{loss}$).
The loaded quality factor ($Q_L$) is determined by the coupling Q of the input and output couplers ($Q_1$ and $Q_2$), the intrinsic Q of the RF cavity ($Q_0$), and the dielectric quality factor ($Q_d$):
\begin{equation}
    \frac{1}{Q_L}=\frac{1}{Q_d}+\frac{1}{Q_1}+\frac{1}{Q_2}+\frac{1}{Q_0}+\frac{1}{Q_d},
\end{equation}
where the dielectric quality factor is inversely proportional to the loss tangent:
\begin{equation}
    p\times\tan\delta=\frac{1}{Q_d}.
\end{equation}

In order to find the optimal parameters of the RF cavity geometry for electro-optic interaction, a trade-off exists between the microwave quality factor, which is reduced by the large dielectric participation ratio, and the maximum electric field in the volume of WGM near the rim of the crystal, which is enhanced by more dielectric participation.
In our design, as the microwave volume is essentially reduced to the volume of the crystal, the losses are mostly concentrated in the dielectric LN resonator, which defines the internal quality factor of the microwave resonator. The resonant frequency is tunable via adjusting the lateral stubs through the quarter wave length. 

\subsection*{Model of the system} 
The electro-optic quantum transduction is based on a three-wave mixing nonlinear process between the optical pump, the optical signal, and the microwave signal. An optical mode $a_p$ at frequency $\omega_p$ is pumped by the coherent light, enabling the coupling and bidirectional conversion between the other optical mode $a$ at frequency $\omega_a$ and a microwave mode ($b$) at $\omega_b$. To satisfy the energy conservation while avoiding the Stokes photon generation into the undesired optical mode, the free spectral range (FSR) of the WGM, which is the frequency difference between two WGMs with consecutive azimuthal mode numbers, is chosen to be slightly larger than the microwave resonance frequency ($\omega_b$). Hence the Hamiltonian of the system is given by \cite{tsang2011cavity}
\begin{equation}
    H = \hbar \omega_{p} a_p^{\dagger} a_p + \hbar \omega_a a^{\dagger} a + \hbar \omega_b b^{\dagger} b - \hbar g_{eo} (b + b^{\dagger}) (a_p+a)^{\dagger} (a_p+a).
    \label{H}
\end{equation}
The single-photon electro-optic coupling strength ($g_{eo}$) is determined by the field overlap between the microwave mode and the WGM:
\begin{equation}
    g_{eo} = n^2r_{33}\sqrt{\frac{\hbar \omega_p \omega_a \omega_b}{8\epsilon_0 \epsilon_b V_p V_a V_b}} \int_{V} \psi_a^*\psi_p \psi_b(r,\theta,\phi) \,dV,
    \label{g0}
\end{equation}
where $n$ is the extraordinary refractive index of LN near 1550nm, $r_{33}$ is the linear electro-optic coefficient; $V_{i}$ ($i=p,a,b$) are the mode volumes of the mode $i$ with dielectric constant $\epsilon_i$, frequency $\omega_i$ and single photon electric field $E_i(r,\theta,\phi)=\sqrt{\hbar\omega_i/2\epsilon_0\epsilon_i V_i}\psi_i(r,\theta,\phi)$. 

As the pump and signal optical modes have a mode number difference of 1, their amplitudes can be expressed as $\psi_{p}=\phi_p(r,\theta) e^{-im\phi}$ and $\psi_{a}=\phi_a(r,\theta) e^{-i(m+1)\phi}$. To match the momentum conservation, a dipole mode is chosen for the microwave cavity whose electric field oscillates along the azimuthal direction as $E_{RF}(r=R_0,\theta=0,\phi)=\widetilde{E}_{RF}(\phi)\cos(\phi)$, where $\widetilde{E}_{RF}(\phi)$ is the correction function that maps the sinusoidal function to the realistic mode distribution. Since the mode size of the WGM is relatively small, we can assume that the microwave electric field has uniform $r$ and $\theta$ distribution inside the WGM volume \cite{hease2020bidirectional}, and therefore calculate $g_{eo}$ as
\begin{equation}
    g_{eo} = \frac{1}{16\pi}n^2r_{33}\sqrt{\omega_p \omega_a} \sqrt{\frac{\hbar\omega_b}{W}}\int_{0}^{2\pi} \widetilde{E}_{RF}(\phi) \,d\phi,
    \label{g}
\end{equation}
where the electric field is normalized by the photon number determined by the total stored microwave energy ($W$) in the RF cavity.
As seen from \refEq{g}, it is essential to maximize the electric field ($\widetilde{E}_{RF}$) by tailoring the microwave mode distribution to obtain a large electro-optic coupling rate. 

To trigger the three-wave mixing process efficiently, a strong coherent pump with frequency $\omega_a-\omega_b$ and detuned from $\omega_p$ by $\delta_p=FSR-\omega_b$ is coupled to the optical resonator. The Hamiltonian in the rotating-wave approximation is reduced to 
\begin{equation}
    H = \hbar \omega_{p} a_p^{\dagger} a_p + \hbar \omega_a a^{\dagger} a + \hbar \omega_b b^{\dagger} b - \hbar g_{eo} (\alpha^*ab^{\dagger} + \alpha a^{\dagger}b),
    \label{H}
\end{equation}
where $\alpha$ is the amplitude of the pump field in the optical cavity.

Assuming an optical driving field $A_{in}e^{-i(\omega_a+\Delta) t}$ and a microwave driving field $B_{in}e^{-i(\omega_b+\delta \Delta) t}$ are applied as the signals for conversion, the dynamics of the system is given by
\begin{equation}
    \frac{da}{dt} = (-i\omega_a-\frac{\gamma_{a}}{2})a+ig\alpha b-\sqrt{\gamma_{a,c}}A_{in}e^{-i(\omega_a+\Delta) t},
\end{equation}
\begin{equation}
    \frac{db}{dt} = (-i\omega_b-\frac{\gamma_{b}}{2})b+ig\alpha^*a-\sqrt{\gamma_{b,c}}B_{in}e^{-i(\omega_b+\Delta) t},
\end{equation}
with
\begin{equation}
    \alpha = \frac{\sqrt{\gamma_{a,c}} A_{in}}{i\delta_p - \frac{\gamma_{a}}{2}},
\end{equation}
where, for a generic $m$ mode ($m=a,b$), $\gamma_{m}$ is the total loss rate, consisting of both the internal loss and external coupling loss: $\gamma_m=\gamma_{m,0}+\kappa_{m,c}$.
By solving the dynamics of the system in the steady state, we can obtain the transfer functions for conversion from the microwave to optical signals, and vice versa, which are equal due to reciprocity. The bi-directional frequency conversion efficiency depends on the  microwave-optical cooperativity and losses:
\begin{equation}
    \eta=\frac{\gamma_{a,c}}{\gamma_a}\frac{\gamma_{b,c}}{\gamma_b}\times\frac{4C}{(1+C)^2}
    ~~,~~
    C=\frac{4n_p{g}_{eo}^2}{\gamma_a\gamma_b}\label{eq_efficiency}
\end{equation}
where $C$ is the cooperativity between the optical and the microwave modes, which depends on the photon number in the pump mode ($n_p$). The second term of Eq.~\ref{eq_efficiency} is defined as the internal efficiency: $\eta_i= 4C / (1+C)^2$, which reaches unity at $C=1$.

As a result, the limiting factors in the attainment of high-efficiency in the microwave-optical transduction are:
\begin{itemize}
    \item The quality factor of the microwave cavity -- While the optical $Q_a$ can reach $10^7$, microwave $Q_b$ is limited in the state-of-art schemes.
    \item Single-photon electro-optic coupling coefficient ($g_{eo}$) -- This paramter is determined by the overlap between microwave and optical fields, and has a trade-off with microwave Q factor. 

    \item Pump power -- High pump power leads to overheating of the device, and thermal quasiparticles poisoning, preventing the device operation in the quantum regime.
\end{itemize}

\subsection*{Simulation of the microwave and optical modes}
\begin{figure}[H] 
	\centering
	\includegraphics[width=1\textwidth]{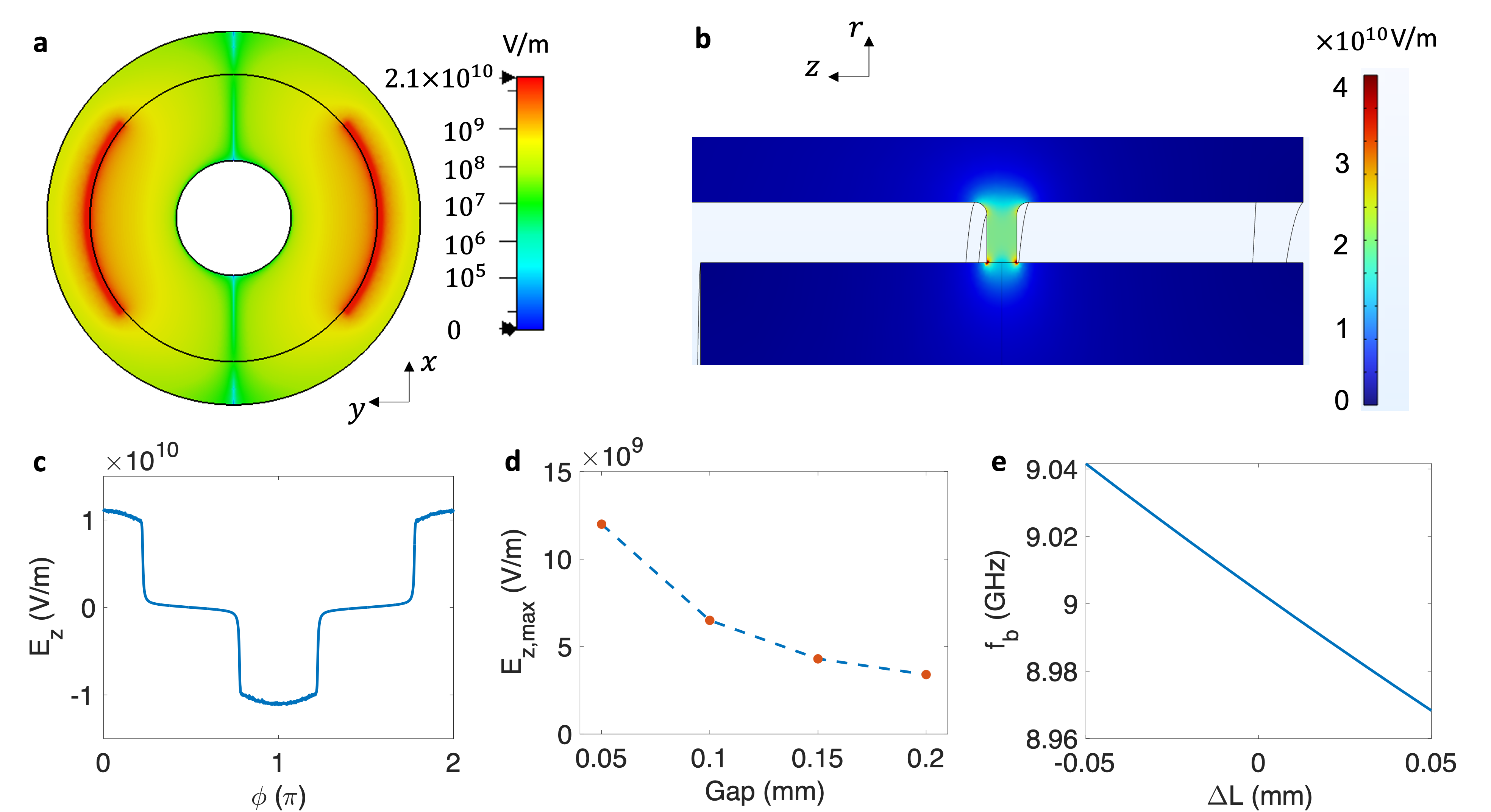}
	\caption{\normalsize Simulation of the microwave mode.  \textbf{a} The z-component of the electric field on the $x-y$ plane of a microwave dipole mode near \SI{9}{\giga\hertz}. \textbf{b} The z-component of the electric field on the $r-\theta$ cross-section plane with $\phi=0$.  \textbf{c} Azimuthal distribution of the z component of the electric field ($E_Z$) along the rim of the LN optical resonator. \textbf{d} Maximum Electric field on the x-y plane shown in \refFig{Fig2}\textbf{a} versus the gap between the LN crystal and the RF cavity wall. \textbf{e} Dependence of the microwave resonance frequency [$f_b=\omega_b/(2\pi)$] on the variation of the quarter wave region length ($\Delta L$). The simulation is based on an stored energy of \SI{1}{\joule}.} 
	\label{Fig2}
\end{figure}

Using Finite-Difference Time-Domain (FDTD) method, we simulate the electric field distribution of a \SI{9}{\giga\hertz} dipole mode in the RF cavity with the LN embedded (\refFig{Fig2}\textbf{a} and \textbf{b}). As noted, the electric field has a strong concentration around the rim of the LN crystal. The dipole mode property is verified by the full oscillation of electric field along the rim of the disk (\refFig{Fig2}\textbf{c}). The field distribution is distorted from a standard sinusoidal shape due to the variation of the gap between the RF cavity wall and the LN crystal. The proper choice of gap size is critical for the optimization of the field magnitude on the rim of the LN crystal (\refFig{Fig2}\textbf{d}). The maximum z component of the electric field is over \SI{1e10}{\volt/\meter}, yielding a single-photon electro-optic coupling coefficient of $2\pi\times$\SI{46.75}{\hertz} based on \refEq{g}. 

As noted before, due to the extremely low niobium loss and high intrinsic Q for the RF cavity, the microwave loss at cryogenic temperature is dominated by the loss tangent of LN crystals. Based on the reported LN loss tangent ($\tan\delta=$\SI{1e-5}{}) \cite{yang2007characteristics}, we estimate the quality factor of the microwave dipole mode to be $1.1\times10^5$ which is well above what obtained in previous works by more than three orders of magnitude. The removal of the central part of the disk ensures lower microwave loss on the redundant LN crystals where WGM does not exist, while further optimization of $Q_b$ is possible by further enlarging the diameter of the central hole, the design is ultimately limited by the fabrication technique. 

As the FSR needs to match with the microwave resonance frequency ($\omega_b$), we study the tunability of $\omega_b$ by applying deformation to the volume of the quarter wave region (\refFig{Fig2}\textbf{e}). The sensitivity of \SI{700}{\mega\hertz/\milli\meter} allows the tunability of the cavity resonant frequency by several tens of MHz and also with very low resolution through piezo stages. 

\begin{figure}[!htb]
	\centering
	\includegraphics[width=1\textwidth]{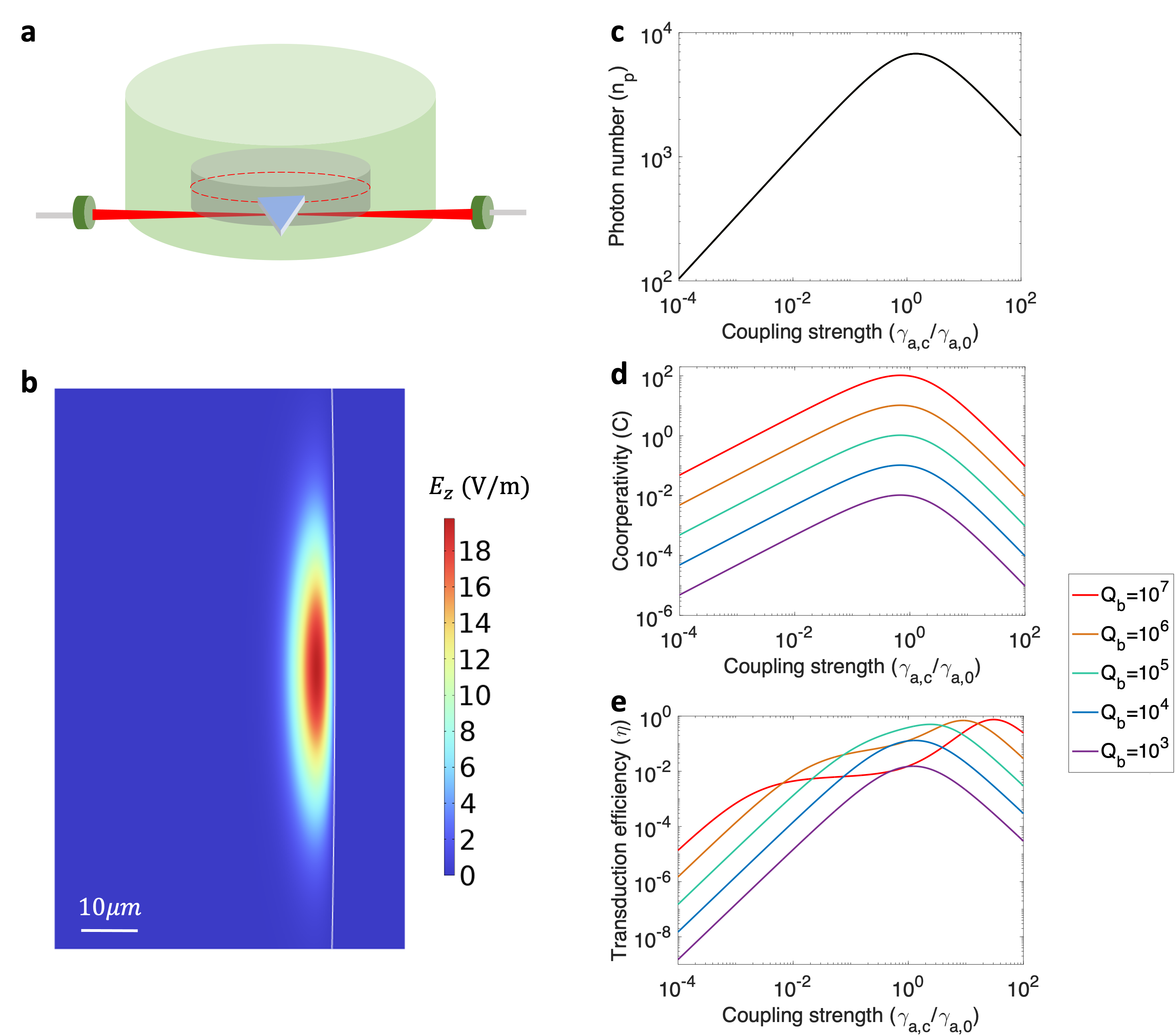}
	\caption{\normalsize Optical design for quantum transduction. \textbf{a} Schematic diagram of the optical coupling scheme with a prism coupler. The optical pump and signal are focused onto the prism coupler via a microlens. The reflected light is collected by another microlens. \textbf{b} Whispering gallery mode in the LN optical resonator. The while lines show the boundary of the LN resonator. \textbf{c} Photon number in the pump mode versus the coupling strength between the prism and the optical resonator ($\gamma_{a,c}$ normalized by the intrinsic optical loss rate $\gamma_{a,0}$). \textbf{d} Cooperativity and \textbf{e} transduction efficiency versus the resonator-prism coupling for different quality factors of the microwave cavity ($Q_b$). Parameters: $Q_a=10^7$, $Q_b=10^5$, $\gamma_{b,c}=3.4\times\gamma_{b,0}$, $g_{eo}=2\pi\times \SI{46.75}{\hertz}$,  $\delta_p=\SI{10}{\mega\hertz}$.}
	\label{Fig3}
\end{figure}

We also simulate the mode distribution of an optical WGM around \SI{192.43}{\tera\hertz} frequency by the FDTD approach (\refFig{Fig3}\textbf{b}). Based on the eigenfrequencies of three WGMs with consecutive mode numbers, we found that the FSR is around \SI{8.93}{\giga\hertz}. The curvature of the side wall is 2.5mm, making it appearing as near-flat. The smooth surface enabled with state-of-art LN fabrication techniques can ensure a high optical quality factor ($Q_a$) above $10^7$. In addition, the upper and bottom surfaces can also be engineered to be slightly rounded depending on fabrication needs. Those small curvatures will have negligible influence on the microwave and optical mode profile. 

In order to couple the optical pump into the WGM, a high-index prism is placed close to the rim of the optical resonator (\refFig{Fig3}\textbf{a}). The coupling strength between the resonator and the prism ($\gamma_{a,c}$) is controlled by the spatial gap between them. It is known that the pump photon number always reaches the maximum at critical coupling \cite{cai2000observation}, i.e., $\gamma_{a,c}$ is equal to the intrinsic loss rate $\gamma_{a,0}$ (\refFig{Fig3}\textbf{c}). 
However, the critical coupling does not lead to the optimized point for transduction efficiency. As shown in \refFig{Fig3}\textbf{d}, the cooperativity is maximized around $\gamma_{a,c}=0.7\times\gamma_{a,0}$ regardless of the microwave quality factor ($Q_b$). The overall transduction efficiency, however, is optimized at different values of coupling losses for different microwave quality factors ($Q_b$) (\refFig{Fig3}\textbf{3}). In our case where $Q_b\sim10^5$, the optimized coupling loss is around $\gamma_{a,c}=2.3\times\gamma_{a,0}$. This is a lightly overcoupled regime that is favorable for efficient readout of the transduction signal. 

\subsection*{Figures of merit of the quantum transducer}
With the device design presented above, we explore the figures of merit of the transduction process at different conditions of operation. In particular, a high-performance transducer should have (1) high coorperativity and efficiency; (2) low operating pump power which indicates low thermal noise; and (3) large bandwidth for signal conversion. 

\begin{figure}[!htb]
	\centering
	\includegraphics[width=1\textwidth]{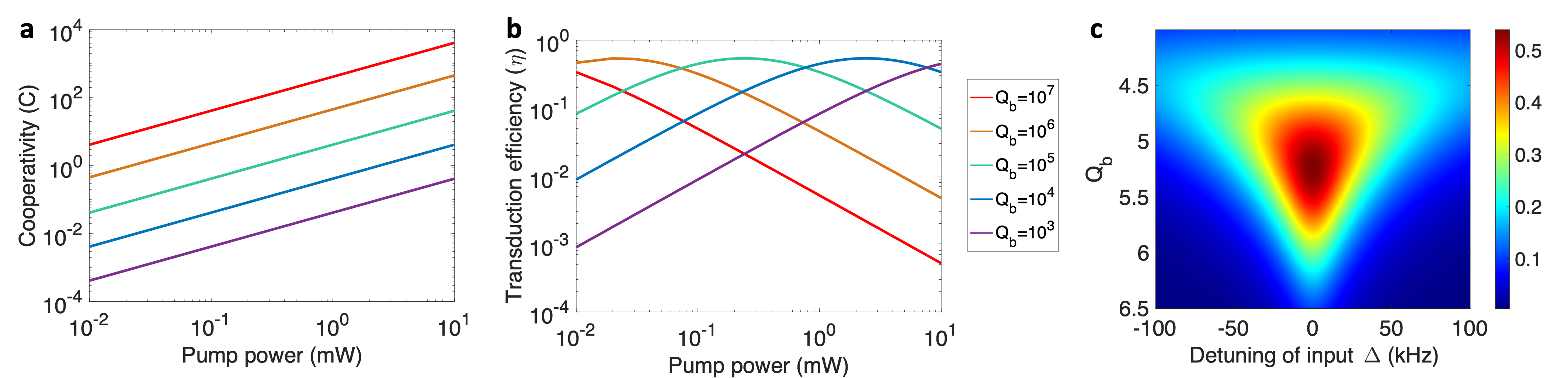}
	\caption{Effect of pump power on transduction. \textbf{a}, \textbf{b} Cooperativity (\textbf{a}) and transduction efficiency (\textbf{b}) as a function of pump power for different quality factors of the microwave cavity ($Q_b$). \textbf{c} Transduction efficiency as a function of $Q_b$  and the frequency detuning of the optical/microwave input ($\Delta_b$). Parameters are the same as in \refFig{Fig3}.}
	\label{Fig4}
\end{figure}

As noted, a high level of optical pump benefits the cooperativity and transduction efficiency, but introduces more overheating of the device that leads to excessive thermal noise photons. Moreover, the stray infrared pump photons that is scattered out of the designed optical path and illuminates on the RF cavity surface can break the cooper pairs of the niobium and deteriorate the coherence of RF cavities and qubits \cite{serniak2018hot,zmuidzinas2012superconducting}. To enable a quantum operation, as shown by previous literature, the pump power needs to be in the sub-mW level \cite{hease2020bidirectional}. Therefore, we study the cooperativity and transduction efficiency as a function of the pump power for our design. We compare the cases with different microwave quality factors, and find out that the conversion efficiency is greatly enhanced by the microwave quality factor ($Q_b$). Our design can achieve 0.58 cooperativity and 50\% conversion efficiency at \SI{140}{\micro\watt} pump power (\refFig{Fig4}\textbf{a} and \textbf{b}), this level of performances greatly exceeds the limit of state-of-art technology. Current transduction designs with lower $Q_b$ are instead limited by the pump power in the attainement of high efficiency.

Moreover, we investigate the bandwidth of the transducer by assuming that the RF cavity is fed by a weak microwave signal with certain frequency detuning ($\Delta$) from the operating frequency $\omega_b$. From simulation, we find out that the bandwidth is enhanced by decreasing $Q_b$ (\refFig{Fig4}\textbf{c}). This is due to the fact that a larger linewidth of the microwave mode can tolerate a wider frequency range of the input signal. Therefore there is always a trade-off between the transduction efficiency and the operating bandwidth. For our design with $Q_b\sim10^5$, the microwave bandwidth ($\Delta_b$) is \SI{100}{\kilo\hertz} with the center of the spectrum window located at $\omega_b$. 

Figures of merits for this design are summarized in the following Table.

\begin{table}[h]
	\centering
	\caption{Main design parameters and figured of merits for the presented 3D transduction device based on the coupling of an SRF cavity and an electro-optic resonator.}\label{CLIC_par_table}
	\begin{tabular}{l|ll|l}
		\hline
		\textbf{Parameter}&\textbf{Symbol}&\textbf{Unit}&\textbf{Value}\\\hline
		Free spectral range&FSR&\si{\giga\hertz}&8.93\\
		Loss tangent & $\tan\delta$ & & 1e-5\\
		\hline
		Optical quality factor & $Q_a$ &  & \SI{1e7}{} \\ 
		Microwave quality factor &$Q_b$ & &\SI{1.1e5}{}\\
		Coupling rate&$g_{eo}$& \si{\hertz} &46.75\\
		Pump frequency detuning&$\delta_p$&\si{\mega\hertz}&10\\Pump power&$P_p$&\si{\micro\W}&140\\
		Cooperativity&C& &0.58\\
		Conversion efficiency&$\eta$&\% &50\\
		Microwave bandwidth&$\Delta_b$ &\si{\kilo\hertz}&100\\
		\hline
	\end{tabular}	
\end{table}

\subsection*{Quantum transduction in a hybrid quantum electrodynamics system}
We now extend the electro-optic quantum transducer to a cavity QED system where the RF cavity is dispersively coupled to a superconducting transmon qubit located within the cavity mode volume. With the hybrid QED system built upon our transducer, one can not only convert the microwave input signal to optical photons, but also optically read out the qubit quantum state via optical measurement. The full Hamiltonian of the hybrid QED system is given as
\begin{equation}
    H = \frac{1}{2}\hbar \omega_{q}^{'}\sigma_z + \hbar \omega_a a^{\dagger} a + \hbar \omega_b^{'} b^{\dagger} b - \hbar g_{eo} (\alpha^* a b^{\dagger}+\alpha a^{\dagger} b)+\hbar \chi b^{\dagger}b\sigma_z,
    \label{H}
\end{equation}
where $\omega^{'}_{q,b}$ refer to the renormalized frequencies of the qubit and the microwave cavity mode, and $\chi$ is the dispersive coupling rate. As indicated by the term $\hbar(\omega_b^{'}+\chi \sigma_z) b^{\dagger} b$, the excited state of the qubit induces a dispersive shift ($\chi$) to the resonance frequency of the RF cavity. Conventionally, the qubit state can be measured by characterizing the frequency shift in the spectrum of the readout cavity. Based on our design, the dispersive frequency shift can be further converted to the amplitude change of the optical signal as the readout of the transducer. Thus, the optical readout and transfer of the state information of the superconducting qubit can be achieved.

\begin{figure}[!htb]
	\centering
	\includegraphics[width=1\textwidth]{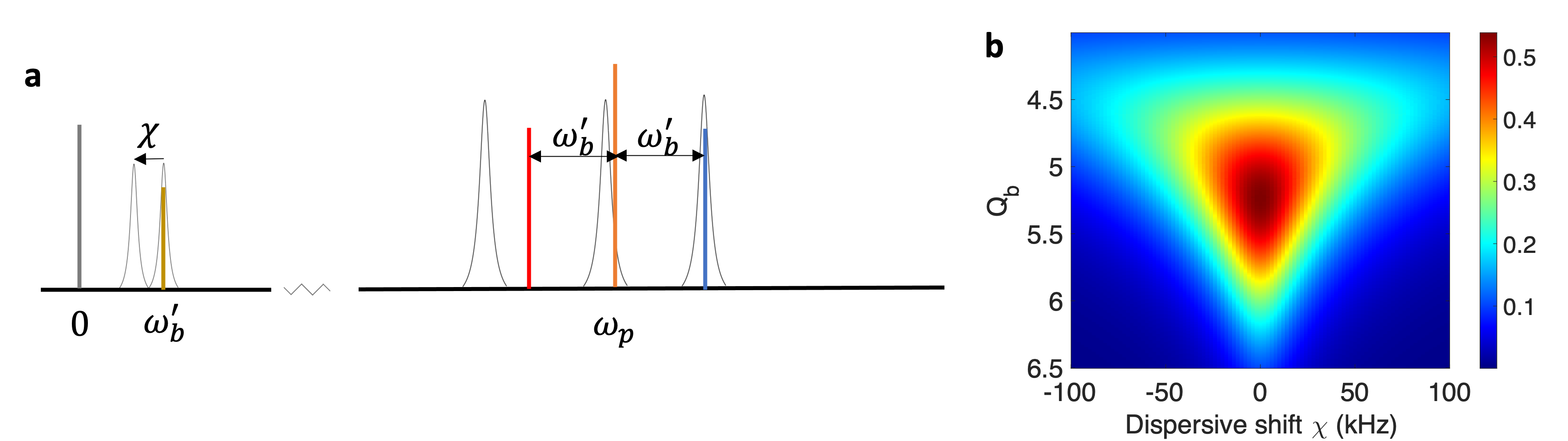}
	\caption{Transduction in a superconducting QED system with a qubit. \textbf{a} Frequency spectrum for the output signal of the quantum transducer.  \textbf{b} Conversion efficiency as a function of the microwave quality factor and the dispersive shift of the microwave resonance frequency induced by the qubit.}
	\label{Fig5}
\end{figure}

We assume that the renormalized RF resonance frequency $\omega^{'}_b$ is equal to $FSR-\delta_p$, and a weak microwave readout pulse with amplitude $B_{in}$ and frequency $\omega=\omega^{'}_b$ is added to the RF cavity. The dynamics of the system in the interaction picture is given by
\begin{equation}
    \frac{d\widetilde{a}}{dt} = (i\chi-\frac{\gamma_a}{2})\widetilde{a}+ig\alpha \widetilde{b},
\end{equation}
\begin{equation}
    \frac{d\widetilde{b}}{dt} = -\frac{\gamma_b}{2}\widetilde{b}+ig\alpha^* \widetilde{a}-\sqrt{\gamma_{b,c}}B_{in}e^{i\chi t},
\end{equation}
where we define $a=\widetilde{a}e^{-i(\omega_a+\chi) t}$ and $b=\widetilde{b}e^{-i(\omega_b^{'}+\chi) t}$. 
The resulting frequency components of the output optical field of the hybrid QED system are shown in \refFig{Fig5}\textbf{a}.

Based on the above analysis and the parameters of our design, we simulate the transduction efficiency, which determines the level of optical output, as a function of the dispersive shift $\chi$ (\refFig{Fig5}\textbf{b}). The transduction efficiency shows quick decay as the magnitude of the dispersive shift increases. This is mainly due to the fact that, when the microwave frequency is detuned from both the optical FSR and the readout pulse, the three-wave mixing process becomes increasingly less efficient. As the $Q_b$ increases, the resolution to distinguish the dispersive shift becomes higher, which can reduce the error rate in the qubit state readout.  

\refFig{Fig4}\textbf{c} and \refFig{Fig5}\textbf{b} show similar behaviors which lead to a trade-off between the bandwidths for input detuning and the resolution of the dispersive shift. At $Q_b=10^5$, we find that our system can tolerate \SI{\pm50}{\kilo\hertz} detuning for the input signal around the RF resonance frequency, while being able to fully resolve \SI{\pm98}{\kilo\hertz} ($\eta<0.1$) dispersive shift induced by the qubit. With typical superconducting qubit-cavity coupling strength up to $\sim$ \SI{}{\mega\hertz} level\cite{koch2007charge}. This scheme can therefore work for a variety of QED systems to optically readout qubit information and connect superconducting qubits in remote quantum processors. 

\subsection*{Heralded entanglement generation based on transducers}
We discuss an application of our high-efficiency quantum transducer to quantum entanglement generation. One of the key challenges for the realization of quantum networks is to create entangled quantum states between two distant quantum nodes. It has been proposed that the optically heralded quantum entanglement can be generated with high-efficiency quantum transducers
\cite{zhong2020proposal,krastanov2021optically}. Such a scheme can work in both the regimes of blue sideband pumping and red sideband pumping, with the latter offering higher fidelity due to its protection against multiphoton generation.

Here, we analyze the entanglement generation features based on the optically heralded approach using our electro-optic transducer. The setup consists of two quantum transducers spatially separated. The optical cavities of the transducers are coupled to optical waveguides, i.e. fiber optics, which send the generated optical photons to a beam splitter followed by two detectors without distinguishing which-path information. The heralded signal indicates the remote generation of entangled microwave photon pairs in superconducting cavities. In entanglement generation, the target state is a Bell state in the two distant microwave cavities $|\psi>=c_1|01>+c_2|10>$. Such a state can be generated by either the blue-sideband pumping or red-sideband pumping approach, as discussed below. 

As for the blue-sideband pumping case, a pump optical photon can be converted into a microwave and an optical photon. The photon generation in one transducer follows a Poissonian process with photon generation rate $r_0$. Therefore, the probability that a photon is generated in each microwave cavity is expressed as 
\begin{equation}
    P_1= r_0 \Delta t e^{-r_0 \Delta t},
\end{equation}
on the other hand, the probability that no photon is generated is 
\begin{equation}
    P_0= e^{-r_0 \Delta t}.
\end{equation}
Therefore, the probability that one photon is generated in each microwave cavity is given as
\begin{equation}
    P_{11}= P_1^2.
\end{equation}
Furthermore, there is a possibility that more than one photon is generated in one microwave cavity. The probability associated with this event is
\begin{equation}
    P_{m>1,n\leq 1}=P_{m\leq 1,n>1}= (1-P_0-P_1)(P_0+P_1).
\end{equation}
Moreover, there can also be more than one photons generated in both cavities, with the probability
\begin{equation}
    P_{m>1,n>1}=(1-P_0-P_1)^2.
\end{equation}
One can write the final state of the system as a superposition of two-photon Fock states:
\begin{align}
\ket{\psi_f}=&\sqrt{P_{00}}\ket{00}+\sqrt{P_{10}}\ket{10}+\sqrt{P_{01}}\ket{01}+\sqrt{P_{11}}\ket{11}+\sqrt{P_{m>1,n\leq 1}}\ket{m>1,n\leq 1}\nonumber\\
&+\sqrt{P_{m\leq 1,n>1}}\ket{m\leq 1,n>1}+ \sqrt{P_{m>1,n>1}}\ket{m>1,n>1},
\end{align}
where we neglect the phase information in the coefficients. Assuming fixed duration time ($\Delta t$) of the transduction process, the entangled microwave photon pair is generated at a rate $2r_0e^{-r_0\Delta t}\frac{\Delta t}{\Delta t+t_r}$, where $t_r$ is the microwave reset time after each generation event \cite{krastanov2021optically}. The infidelity corresponds to the probability that each microwave cavity contains one photon, or that more than one photon is generated in at least one cavity. Therefore the fidelity is expressed as
\begin{equation}
\eta_b=\frac{P_{10}+P_{01}}{P_{10}+P_{01}+P_{11}+P_{m>1,n\leq 1}+P_{m\leq 1,n>1}+P_{m>1,n>1}}.
\end{equation}

In the red-sideband pumping case, the remote entanglement is enabled by first preparing both microwave cavities at the one-photon state while leaving the optical cavity at the zero-photon state. Upon the simultaneous onset of the transduction processes in both transducers, a microwave photon can be converted to an optical photon that triggers the "click" of the detector, with the probability
\begin{equation}
    P_0= 1-\exp(-r_0 \Delta t).
\end{equation}
As such, the probability that no optical photon is generated is rather given by
\begin{equation}
    P_1= \exp(-r_0 \Delta t).
\end{equation}
As there cannot be more than one photon generated in a microwave cavity, the generated state of the whole system including two quantum units is simply $\ket{\psi_f}=\sqrt{P_{00}}\ket{00}+\sqrt{P_{10}}\ket{10}+\sqrt{P_{01}}\ket{01}+\sqrt{P_{11}}\ket{11}$. Moreover, the infidelity is only affected by $P_{00}=P_{0}^2$, which leads to
\begin{equation}
\eta_r=\frac{P_{10}+P_{01}}{P_{00}+P_{10}+P_{01}}.
\end{equation}

\begin{figure}[!htb]
	\centering
	\includegraphics[width=1\textwidth]{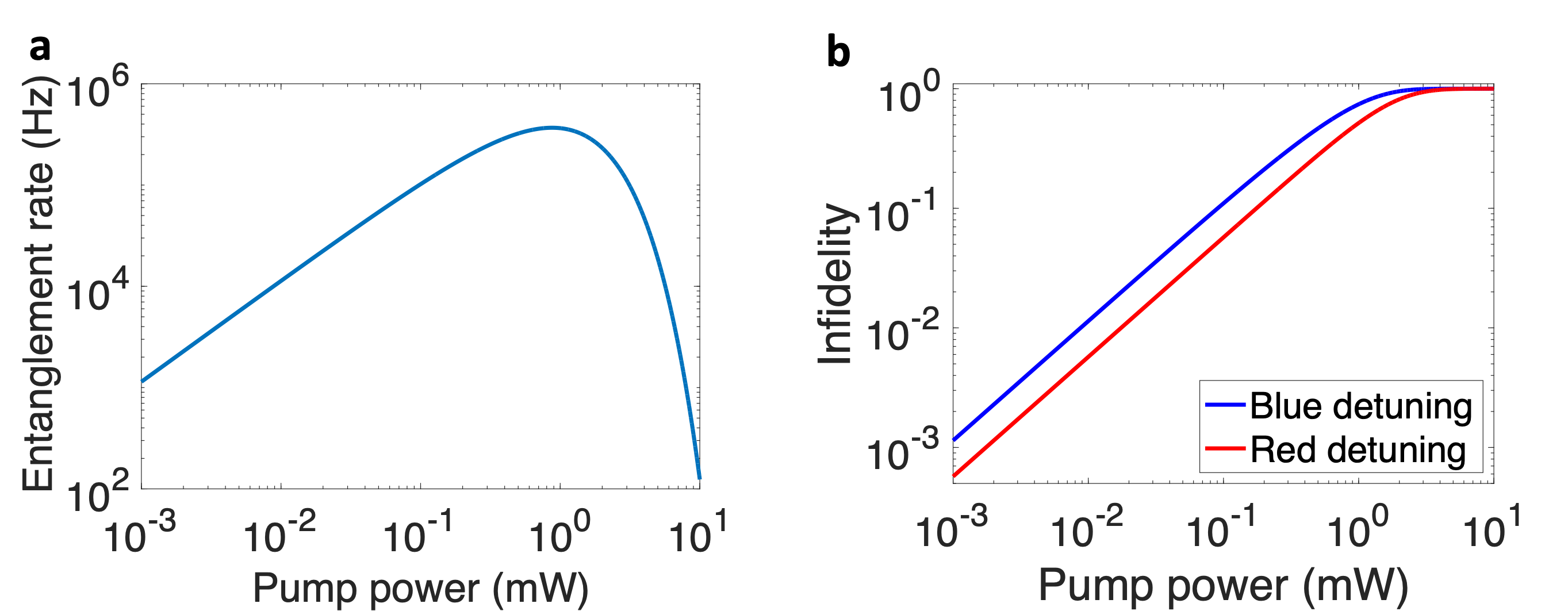}
	\caption{Optically heralded entanglement generation based on electro-optic quantum transducers. \textbf{a}, \textbf{b} Entanglement generation rate (\textbf{a}) and infidelity (\textbf{b}) as a function of pump power. Microwave reset time $t_r=1\mu s$. Duration time $\Delta t=1\mu s$.}
	\label{Fig6}
\end{figure}

Here, we analyze the rate and infidelity of entanglement generation based on the parameters in our design of the transducers. While the entanglement generation rate is optimized at a pump power higher than \SI{1}{\milli\watt}, the increase of pump power reduces the fidelity, due to additional possibility of generating more than one microwave photons (\refFig{Fig6}). This trade-off limits the amount of pump power that can be applied, as well as the speed of entangled states generation for distributed quantum communication. For our design, a \SI{20}{\kilo\hertz} entanglement rate can be realized at a pump power of \SI{18}{\micro\watt} with an infidelity about 0.01.

\subsection*{Enhanced quantum sensing with high-coherence transduction devices}
The high-efficiency microwave-optical transducer can also be utilized to enhance the measurement precision of weak microwave signals in quantum sensors. The microwave signal measurement is of critical importance in various fundamental studies and applications, including, for instance, single photon detection, highly sensitive axion and dark photon haloscope measurement, new physics particle searches in the \si{\tera\hertz} range and so on \cite{alam2022quantum}. The up-conversion operation performed with this hybrid transduction system, the microwave information is converted to the optical regime through self-heterodyne techniques, which have the advantage of eliminating the local oscillator (LO) used in conventional super-heterodyne mixers, reducing therefore complexity, power consumption, added noise, and thermal quasiparticle poisoning.
One issue associated with the transducer working for precise measurement is that the conversion process introduces additional single-photon shot noise and back-action noise. The standard noise spectrum of microwave detection using single-quadrature measurement is given by:
\begin{equation}
    S(\Delta)=2\kappa_b (2n_T+1)+\frac{(\kappa_b^2+\Delta^2)(\kappa_a^2+\Delta^2)}{4n_p g_{eo}^2} + \frac{4n_p g_{eo}^2}{\kappa_a^2+\Delta^2},
\end{equation}
which is ultimately limited by both the standard quantum limit (SQL) and the RF noise floor as $S(\Delta)\geq S_{RF}+S_{SQL}$, where $S_{RF}=2\kappa_b(2n_T+1)$ and $S_{SQL}=\sqrt{\kappa_b^2+\Delta^2}$.

\begin{figure}[!htp]
\centering
 \includegraphics[width=0.6\textwidth]{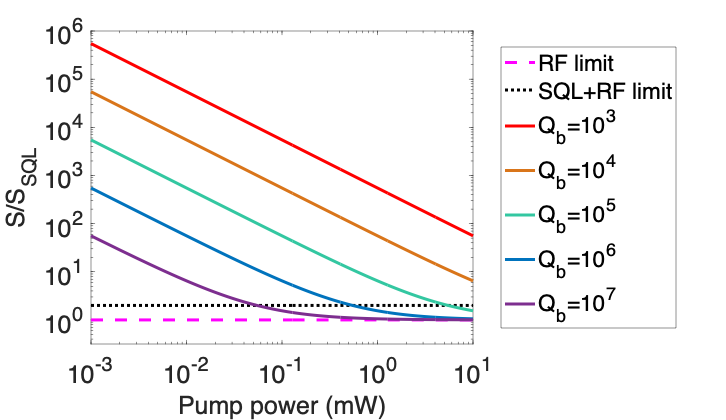}
    \caption{High-efficiency transducer for high-precision microwave measurement. Noise spectral density (S) normalized by the standard quantum limit (SQL) as a function of the pump power, with back-action noise cancellation. The parameters follow our transducer design, except that the optical Q factor is chosen to be $10^8$.} \label{Fig7}
\end{figure}

The enhancement of microwave photons measurements at the quantum threshold through transduction methods and with the assistance of back-action noise cancellation techniques have been reported in Ref.~\cite{nazmiev2022back}. In particular, a judicious choice of combination of independent quadrature measurements can lead to the cancellation of the back-action noise in the transduction process \cite{nazmiev2022back}. In the back-action evading approach, the detection noise density is rather given by:
\begin{equation}
    S(\Delta)=2\kappa_b (2n_T+1)+\frac{(\kappa_b^2+\Delta^2)(\kappa_a^2+\Delta^2)}{C\kappa_a \kappa_b},
\end{equation}
where $n_T$ is the number of thermally excited photons and $C$ is the cooperativity for transduction following \refEq{eq_efficiency}.
Therefore, with the back-action scheme, there is a possibility to break the standard detection limit with reference to the SQL. Fig.~\ref{Fig7} shows noise spectral densities as a function of pump power for different microwave Q's. The noise level of the sensor can be lowered down to the RF limit breaking the SQL limit by increasing the microwave coherence time or increasing the pump power. Particularly, the large cavity coherence can significantly reduce the required pump power to obtain a high precision for detection, making it meaningful to further increase the coherence time of cavities. 

\section*{\bf Discussion}
We design a three-dimensional electro-optic system for high-efficiency and low-noise microwave-optical quantum transduction based on the SRF technology developed at Fermilab. The transduction process is optimized by calculating the figures of merit versus various degrees of freedom in the SRF cavity geometry, LN crystal geometry, optical coupling, and optical pump power.  
Through microwave and optical analysis and simulations we show transduction performance and up to 50\% frequency conversion at very low pump power. Compared to the previous, this design breaks the usual trade-off between transduction efficiency and quantum noise.

This platform holds great potential for near-future realization of several applications. The electro-optic quantum transducer allows for optical readout of quantum information in superconducting QED systems inside the dilution fridges \cite{lecocq2021control}. It provides a critical interface for the achievement of huge scaling capabilities for quantum computers, going beyond the limit of local dilution refrigerators, building links between superconducting quantum processors, and realizing remote entanglement over optical fibers.

High conversion efficiency can also be leveraged in quantum sensing techniques breaking the standard quantum limit (SQL) by applying back action evading schemes or squeezing techniques for the detection of microwave photons. Thus, these transduction techniques can serve as a platform for fundamental physics experiments such as dark matter detection over a wide frequency range.

\section*{\bf Data Availability}
The numerical data generated in this work is available from the authors upon reasonable request.

\section*{\bf Acknowledgment}
This manuscript has been authored by Fermi Research Alliance, LLC under Contract No. DE-AC02-07CH11359 with the U.S. Department of Energy, Office of Science, Office of High Energy Physics. This work is funded by the Fermilab’s Laboratory Directed Research and Development (LDRD) program.

This research used resources of the U.S. Department of Energy, Office of Science, National Quantum Information Science Research Centers, Superconducting Quantum Materials and Systems Center (SQMS) under contract number DE-AC02-07CH11359. The NQI Research Center SQMS contributed by supporting the design of SRF cavities and access to facilities. 

The authors would like to thank Johannes Fink for the insightful discussions and feedback on topics related to this paper.

\section*{Author Contributions}

C.W. carried out numerical analysis and  optical simulations, coordinated the the microwave-optical design, and drafted the manuscript.  S.Z. conceived the project, coordinated the research and helped drafting and editing the manuscript. I.G. conducted microwave simulations and analysis. S.K. performed calculations and along with V.Y. contributed to the conceptual design. A.G. and A.R. contributed to the final version of the manuscript. All authors gave their final approval for publication.

\section*{\bf Competing Interests}
The authors declare no competing interests. 

\bibliographystyle{unsrt}

\end{document}